\documentclass[aps,prd,twocolumn,10pt,floatfix,amsfonts,amssymb,amsmath,nofootinbib,superscriptaddress]{revtex4-1}
\pdfoutput=1

\usepackage[utf8]{inputenc}
\usepackage{lmodern}
\usepackage[T1]{fontenc}
\usepackage{xcolor}
\usepackage{graphicx}
\usepackage{ifpdf}
\usepackage{units}
\usepackage{suffix}
\usepackage{booktabs}
\ifpdf
\else
  \PassOptionsToPackage{dvipdfmx}{hyperref}
\fi

\usepackage{hyperref}
\usepackage{textcomp}
\usepackage{etoolbox}

\usepackage{nicefrac}
\usepackage[h]{esvect}
\let\vec\vv

\newcommand{\dd}[2][]{\ifstrempty{#1}{\mathop{\mathrm{d}#2}}{\mathop{\mathrm{d}^{#1}#2}}}
\newcommand{\I}{\mathrm{i}}
\DeclareMathOperator{\Div}{\vv{\nabla}\cdot}
\DeclareMathOperator{\rot}{\vv{\nabla}\times}
\DeclareMathOperator{\grad}{\vv{\nabla}}
\DeclareMathOperator{\sgn}{sgn}

\graphicspath{{figures/}}

\newcommand{\Fig}[1]{Fig.~\ref{#1}}
\newcommand{\Sec}[1]{Sec.~\ref{#1}}
\newcommand{\Eq}[1]{Eq.~\eqref{#1}}
\newcommand{\Eqs}[3][-]{Eqs.~\eqref{#2}#1\eqref{#3}}

\begin{document}

\title{Efficient retrieval of phase information from real-valued electromagnetic field data}

\author{Alexander \surname{Blinne}}
\email{alexander.blinne@uni-jena.de}
\affiliation{Helmholtz Institute Jena, Fr\"obelstieg 3, D-07743 Jena, Germany}

\author{Stephan \surname{Kuschel}}
\affiliation{Helmholtz Institute Jena, Fr\"obelstieg 3, D-07743 Jena, Germany}
\affiliation{Institute for Optics- and Quantum Electronics, Friedrich Schiller University Jena, Max-Wien-Platz 1, 07743 Jena, Germany}

\author{Stefan \surname{Tietze}}
\affiliation{Helmholtz Institute Jena, Fr\"obelstieg 3, D-07743 Jena, Germany}

\author{Matt \surname{Zepf}}
\affiliation{Helmholtz Institute Jena, Fr\"obelstieg 3, D-07743 Jena, Germany}
\affiliation{Institute for Optics- and Quantum Electronics, Friedrich Schiller University Jena, Max-Wien-Platz 1, 07743 Jena, Germany}
\affiliation{Department of Physics and Astronomy, Queen's University Belfast, Belfast, BT7 1NN, UK}

\begin{abstract}
While analytic calculations may give access to complex-valued electromagnetic field data which allow trivial access to envelope and phase information,
the majority of numeric codes uses a real-valued represantation.
This typically increases the performance and reduces the memory footprint, albeit at a price:
In the real-valued case it is much more difficult to extract envelope and phase information, even more so if counter propagating waves are spatially superposed.
A novel method for the analysis of real-valued electromagnetic field data is presented in this paper.
We show that, by combining the real-valued electric and magnetic field at a single point in time, we can directly reconstruct the full information of the electromagnetic fields in the form of complex-valued spectral coefficients ($\vv{k}$-space) at a low computational cost of only three Fourier transforms.
The method allows for counter propagating plane waves to be accurately distinguished as well as their complex spectral coefficients, i.\,e. spectral amplitudes and spectral phase to be calculated.
From these amplitudes, the complex-valued electromagnetic fields and also the complex-valued vector potential can be calculated from which information about spatiotemporal phase and amplitude is readily available.
Additionally, the complex fields allow for efficient vacuum propagation allowing to calculate far field data or boundary input data from near field data.
An implementation of the new method is available as part of PostPic\footnote{See \cite{PostPicPaper,PostPicGitHub}}, a data analysis toolkit written in the Python programming language.
\end{abstract}

\maketitle

\section{Introduction}
Whenever the physics of some phenomenon or technique becomes so complex, that it cannot be handled using pen and paper or even numerical solutions of analytic equations, physicists will turn towards simulations.
This is true across most sections of physics, including, but not limited to, astrophysics, hydrodynamics, laser physics, plasma physics, material sciences or even quantum physics.
Many of these simulations include charged particles, surfaces or bulk materials, that interact with electromagnetic fields.
Therefore most simulations use the electromagnetic fields $E_{x,y,z}$ and $B_{x,y,z}$ as real variables on a grid.
This especially includes Particle in Cell (PIC) simulations \cite{Yee1966,ArberEPOCH} which are being heavily used to model high power laser-plasma interactions.
While looking at any of these fields directly may already give a lot of insight on what is happening during the simulation, much more information can be extracted when the full physical meaning of the electromagnetic field as a whole is taken into account.
The simplest example for the additional information that can be gained access to is the differentiation between incident and outgoing waves, even when they are superposed and counterpropagating.
Furthermore our analysis gives direct access to the complex field data, i.\,e. envelope and phase of the propagating waves.
This includes the complex fields $\vv{E}$ and $\vv{B}$, as well as the vector potential $\vv{A}$.

The available techniques that give access to the full complex $\vec{k}$-space have some drawbacks which explains why they are not yet in widespread use.
One approach considers the fields in two consecutive dumps, which gives access to the temporal derivative of the fields.
Another approach is based on considering the fields only at some plane, but at equal temporal intervals throughout the simulation.
Either approach requires to write a vast amount of data to disk, which slows down the simulation and is often impractical due to time and/or memory constrains.

The method presented in this work has the advantage that it can reconstruct the full complex $\vv{k}$-space from the information contained in a single standard data dump which contains only the real-valued electric and magnetic fields.
As this method is based upon a spectral representation, the heavy lifting of the computation is done by a single discrete fourier transform for each input field component.
Afterwards the reconstructed fields can be directly calculated, without the need to solve equations numerically or using an iterative method.
Additionally, very fast implementations are readily available for discrete fourier transforms.
Apart from the high computational efficiency, we managed to adapt this method to some of the specific characteristics of finite difference time domain solvers which greatly improves the accuracy as compared to a naïve approach that is based solely on Maxwell theory.

In the one-dimensional case it is well known, that the waves propagating in either direction can be separated using information from only a single simulation dump by the linear combinations
\begin{align}
 \label{eqn:separation_1d_known}
 E_y^+ &= \frac12 \left( E_y+cB_z \right)\,, & E_y^- &= \frac12 \left( E_y-cB_z \right)\,,
\end{align}
which is true in both, spatial domain and frequency domain.

We show, that it is possible, to extend this method into the three dimensional frequency domain and separate the counterpropagating waves in the data using just a single dump.
By separating the waves propagating in the $\vec{k}$ and $-\vec{k}$ direction, we reconstruct the complex amplitudes of any physical mode of the electromagnetic field.
This is different from a simple Fourier transform of the real data where the amplitudes for $\vec{k}$ and $-\vec{k}$ would be Hermitian conjugates making both propagation directions indistinguishable. This is not the case in our reconstructed, complex $\vec{k}$-space.
Here, the amplitudes for $\vec{k}$ and $-\vec{k}$ are completely independent and refer directly to plane waves propagating in either direction with known amplitude and phase.
An inverse Fourier transform of the complex $\vec{k}$-space will then yield a complex field in spatial domain whose real part is, up to round-off errors, identical to the real field that the process was started with.
From these complex spatial data it is easy to access the envelope and spatial phase of the waveform.

By applying an additional phase $\Delta\varphi=e^{-\I \omega(\vec{k}) t}$ to the complex $\vec{k}$-space data, it is trivial to perform additional vacuum propagation of the field.
This is useful to gain access to far field information which spans a bridge to experimental data from, e.\,g., wavefront sensors.

It is also possible to use this method to calculate numerical boundary input data such that, in a simulation, a specific focus shape on target is achieved.
In order to do this, our method is applied not to a simulation dump, but to a model of the fields in the desired laser focus.
This is especially useful if analytic solutions for the propagation of fields of the desired focus are not available or complicated.

\section{Electromagnetic plane waves}
\label{sec:theory}
The electromagnetic fields $\vec{E}(\vec{r},t)$ and $\vec{B}(\vec{r},t)$ can be defined in terms of the vector potential $\vec{A}(\vec{r},t)$ and a scalar potential $\varphi(\vec{r},t)$.
The fields can be calculated from these potentials using the well known relations
\begin{align}
 \vv{E} &= - \grad \varphi - \frac{\partial \vv{A}}{\partial t}\,,
 &
 \vv{B} &= \rot \vv{A}\,,
\end{align}
which guarantee
\begin{align}
 \rot \vv{E} &= -\frac{\partial \vv{B}}{\partial t}\,,
 &
 \Div \vv{B} &= 0\,.
\end{align}
In the absence of matter $\rho=0$, $\vec{j}=\vec{0}$ and chosing radiation gauge $\Div \vv{A}=\varphi=0$, all that is left of Maxwell's equations is the wave equation
\begin{align}
\label{eqn:wave}
 \Box \vv{A} = \left( \Delta - \frac{1}{c^2}\frac{\partial}{\partial t} \right)\vv{A}=0\,.
\end{align}

Introducing the wave vector $\vv{k} = k\hat{k}$, with $|\hat{k}| = 1$, which specifies the direction of propagation $\hat{k}$ of a plane wave with wavelength $\frac{2\pi}{k}$, \Eq{eqn:wave} is solved by the ansatz
\begin{align}
 \label{eqn:ansatz-a}
 \vv{A}(\vv{r},t) &= \int \dd[3]{k} \vv{\mathbf{A}}_{\vv{k}}  e^{\I\left(\vv{k}\cdot\vv{r}-\omega(\vv{k})(t-t_0)\right)}\,,
\\
 \label{eqn:mode_a_canonical}
 \vv{\mathbf{A}}_{\vv{k}} &= \mathbf{a}_1(\vv{k}) \hat{e}_1(\hat{k}) + \mathbf{a}_2(\vv{k}) \hat{e}_2(\hat{k})\,.
\end{align}
Here we have introduced an orthonormal basis $\hat{e}_1, \hat{e}_2$, spanning the plane perpendicular to the wave vector, such that the vectors $\hat{k}, \hat{e}_1, \hat{e}_2$ form a right-handed, orthonormal basis of $\mathbb{R}^3$.
This ansatz is commonly used to quantize the photon field \cite{Greiner1996}.

In order to solve \Eq{eqn:wave},  $\omega(\vv{k})$ needs to be fixed as $\omega(\vv{k})=\omega_\mathrm{vac.}(\vv{k})=ck$.
However, the discretization of a simulation may cause a different dispersion relation $\omega(k)=\omega_\mathrm{grid}(\vv{k})$\cite{Yee1966,Pukhov1999,Cowan2013,Lehe2013,Blinne2017} to be at work.
In order to produce an accurate result from the simulation data, it might be necessary to insert the grid dispersion relation into our ansatz instead of $\omega_\mathrm{vac.}(\vv{k})$.
We will show in \Sec{sec:gaussian_pulse}, that the simple case $\omega(\vv{k})=ck$ is already the correct choice for our example.

In this representation the amplitudes $\mathbf{a}_i(\vv{k}),\,i=1,2$ of the the vector potential are fixed at an arbitrary time $t_0$ and as such do not themselves depend on the time $t$.
These amplitudes contain the full information of the electromagnetic field and any choice of these amplitudes is a solution to Maxwell's equations.

Calculating the electromagnetic fields from the ansatz yields the representations
\begin{align}
 \label{eqn:ansatz_e_canonical}
 \vv{E}(\vec{r},t)   &=    \int \dd[3]{k} \vv{\mathbf{E}}_{\vv{k}}   e^{\I\left(\vv{k}\cdot\vv{r}-\omega(\vv{k})(t-t_0)\right)}\,,
 \\
 \label{eqn:ansatz_b_canonical}
 \vv{B}(\vec{r},t)   &=    \int \dd[3]{k} \vv{\mathbf{B}}_{\vv{k}}   e^{\I\left(\vv{k}\cdot\vv{r}-\omega(\vv{k})(t-t_0)\right)}\,,
\end{align}
with
\begin{align}
 \label{eqn:mode_e}
 \vv{\mathbf{E}}_{\vv{k}}  &= \I\omega(\vv{k}) \vv{\mathbf{A}}_{\vv{k}} \\
 \label{eqn:mode_e_canonical}
&=  \I\omega(\vv{k}) \left(  \hat{e}_1(\hat{k}) \mathbf{a}_1(\vv{k}) + \hat{e}_2(\hat{k}) \mathbf{a}_2(\vv{k}) \right)\,,
 \\
 \label{eqn:mode_b_canonical}
 \vv{\mathbf{B}}_{\vv{k}} &=  \I k  \left( \hat{e}_2(\hat{k}) \mathbf{a}_1(\vv{k}) -  \hat{e}_1(\hat{k}) \mathbf{a}_2(\vv{k}) \right)\,.
\end{align}
Here, the electromagnetic fields and associated amplitudes are considered to be complex.
The linearity of Maxwell's equations in vacuum implies that they are fulfilled by the real and imaginary parts of the complex fields, individually.

Each of the complex fields $\vec{A}(\vec{r},t)$, $\vec{E}(\vec{r},t)$ and $\vec{B}(\vec{r},t)$, individually and at any time $t$, contains the full electromagnetic information, as \Eqs[ and ]{eqn:mode_e_canonical}{eqn:mode_b_canonical} can be easily inverted to find $\mathbf{a}_i(\vv{k})$ for $i=1,2$.
We will show how the complex information can be reconstructed, when only the real parts $\Re(\vec{E})$ and $\Re(\vec{B})$ are available.

Using \Eqs[ and ]{eqn:mode_e_canonical}{eqn:mode_b_canonical} the spectral representations of the gauge condition and of Maxwell's equations
\begin{align}
 \label{eqn:gauge_spectral}
 \vec{k}\cdot\vv{\mathbf{A}}_{\vv{k}} & = 0\,,
\\
\label{eqn:maxwell_gauss}
 \vec{k}\cdot\vv{\mathbf{E}}_{\vv{k}} & = 0\,,
\\
\label{eqn:maxwell_gauss_b}
 \vec{k}\cdot\vv{\mathbf{B}}_{\vv{k}} & = 0\,,
\\
 \label{eqn:maxwell_ampere_eqn}
 \vec{k}\times\vv{\mathbf{B}}_{\vv{k}}  & = \frac{-k^2}{\omega(\vec{k})} \vv{\mathbf{E}}_{\vv{k}}\,,
\\
  \label{eqn:maxwell_faraday_eqn}
 \vec{k}\times\vv{\mathbf{E}}_{\vv{k}} & = \omega(\vec{k}) \vv{\mathbf{B}}_{\vv{k}}
\end{align}
can be straightforwardly found, which will be used in the following to reconstruct the complex $\vv{k}$-space.

\subsection{Fourier Transforms}
Consider this pair of backward and forward Fourier transforms
\begin{align}
 \label{eqn:fourier_backward}
 F(\vec{r}) &= \int\dd[3]{k} e^{\I \vec{k}\cdot\vec{r}} F^\mathrm{F}(\vec{k})\,,
\\
\label{eqn:fourier_forward}
F^\mathrm{F}(\vec{k}) &= \int\mathop{\frac{\dd[3]{r}}{(2\pi)^3}} e^{-\I \vec{k}\cdot\vec{r}} F(\vec{r})\,.
\end{align}
If the function $F(\vec{r})$ is real-valued, it is a well-known fact that the corresponding amplitudes will obey a Hermitian symmetry
\begin{align}
 F^\mathrm{F}(\vec{k}) = F^{\mathrm{F}*}(-\vec{k})\,,
\end{align}
where the $^*$ denotes complex conjugation.
This is the manifestation of the inabilty to distinguish forward and backward propagating waves in a real-valued snapshot of an individual field.

Let us now assume, that the electromagnetic fields in question obey the representation introduced in \Eq{eqn:ansatz-a}, but only their real parts $\Re\left[ \vec{E}(\vec{r}, t_0) \right]$ and $\Re\left[\vec{B}(\vec{r}, t_0)\right]$ are known.
This simulates the situation of being presented with real-valued data from a simulation, under the assumption that the simulation solves Maxwell's equations.
Inserting the real part of \Eq{eqn:ansatz_e_canonical} into \Eq{eqn:fourier_forward} will yield the result of a Fourier transform of the real-valued simulation data, in terms of the spectral coefficients of the complex fields which we want to reconstruct.

We find
\begin{align}
 \vv{E}^\mathrm{F}(\vv{k})
 &=  \int\mathop{\frac{\dd[3]{r}}{(2\pi)^3}} e^{-\I \vec{k}\cdot\vec{r}}\Re\left[ \vec{E}(\vec{r}, t_0) \right]
\\
\label{eqn:fourier_amplitude_e}
&= \frac12 \left( \vv{\mathbf{E}}_{\vv{k}} + \vv{\mathbf{E}}^*_{-\vv{k}}\right)\,,
\end{align}
where the identity $\int \dd[3]{r} e^{\I\vec{r}\cdot(\vv{k}-\vv{k}')} = (2\pi)^3 \delta(\vv{k}-\vv{k}')$ for the Dirac delta distribution has been used.
The analogous result for the magnetic field reads
\begin{align}
\label{eqn:fourier_amplitude_b}
\vv{{B}}^\mathrm{F}(\vv{k})
&= \frac12 \left( \vv{\mathbf{B}}_{\vv{k}} + \vv{\mathbf{B}}^*_{-\vv{k}}\right)\,.
\end{align}

Starting from \Eq{eqn:fourier_amplitude_b} and using \Eq{eqn:maxwell_ampere_eqn} it is straightforward to find
\begin{align}
\label{eqn:fourier_amplitude_kXb}
 \vv{k}\times\vv{{B}}^\mathrm{F}(\vv{k})
&=\frac{k^2}{2\omega(\vv{k})}\left( \vv{\mathbf{E}}^*_{-\vv{k}} - \vv{\mathbf{E}}_{\vv{k}} \right)\,,
\end{align}
such that we can combine \Eqs[ and ]{eqn:fourier_amplitude_e}{eqn:fourier_amplitude_kXb} to find
\begin{align}
\label{eqn:kspace_final_e}
   \vv{\mathbf{E}}_{\vv{k}}
&=\vv{{E}}^\mathrm{F}({\vv{k}}) - \frac{\omega(\vv{k})}{k^2} \vv{k}\times\vv{{B}}^\mathrm{F}({\vv{k}})\,.
\end{align}
In this result we have successfully reconstructed the full information about the electromagnetic field in terms of the complex amplitudes $\vv{\mathbf{E}}_{\vv{k}}$ of the electric field, just from a single snapshot of the respective real parts of the electric and magnetic fields.
Finally inserting $\omega(\vv{k})=\omega_\mathrm{vac.}(\vv{k})=ck$ into \Eq{eqn:kspace_final_e} yields
\begin{align}
   \label{eqn:kspace_final2_e}
   \vv{\mathbf{E}}_{\vv{k}} &= \vv{{E}}^\mathrm{F}({\vv{k}})- c\hat{k}\times\vv{{B}}^\mathrm{F}({\vv{k}})\,.
\end{align}
According to this result the complex $\vv{k}$-space can be calculated by taking the Fourier transform of the real-valued fields as obtained from the simulation $\vv{{E}}^\mathrm{F}({\vv{k}})$ and $\vv{{B}}^\mathrm{F}({\vv{k}})$ and combining them to obtain $\vv{\mathbf{E}}_{\vv{k}}$.

Following the same procedure, but reversing the roles of the electric and magnetic fields, and employing the Maxwell-Faraday equation \Eq{eqn:maxwell_faraday_eqn} instead of the Maxwell-Ampere equation \Eq{eqn:maxwell_ampere_eqn}, we find the analoge result for the complex amplitudes of the magnetic field,
\begin{align}
 \label{eqn:kspace_final2_b}
   \vv{\mathbf{B}}_{\vv{k}} &= \vv{{B}}^\mathrm{F}({\vv{k}})  + \frac{1}{c} \hat{k}\times\vv{{E}}^\mathrm{F}({\vv{k}})\,.
\end{align}

Additionally, by inverting \Eq{eqn:mode_e}, we can also calculate the complex amplitudes of the vector potential
\begin{align}
 \label{eqn:kspace_final2_a}
  \vv{\mathbf{A}}_{\vv{k}} &= \frac{1}{\I\omega(\vv{k})}\vv{\mathbf{E}}_{\vv{k}}\,.
\end{align}
Furthermore, by inverting \Eq{eqn:mode_a_canonical} we find the spectral coefficients
\begin{align}
\label{eqn:spectral_coefficients_a}
 \mathbf{a}_i(\vv{k}) &= \hat{e}_i \cdot \vv{\mathbf{A}}_{\vv{k}},\quad i=1,2\,,
\end{align}
proving that we indeed have reconstructed the full information about the electromagnetic fields.

By applying an inverse Fourier transform according to \Eqs[, ]{eqn:ansatz-a}{eqn:ansatz_e_canonical} or \eqref{eqn:ansatz_b_canonical} to the respective complex amplitudes, the complex fields in spatial domain can be calculated at any point in time.

Our result \Eq{eqn:kspace_final2_e} is similar to a projection operator defined in \cite{Kolesik2002}.
The difference is that, instead of projecting to two distinct subspaces with a specific propagation axis, we combine both parts to a complex field which contains all the information at once, for all directions of propagation.

Reducing \Eq{eqn:kspace_final2_e} to one dimension via inserting $k_y=k_z=0$, we find
\begin{align}
   \label{eqn:kspace_final_e1d}
   \mathbf{E}_{y}(k_x) &= \vv{E}^\mathrm{F}_{y}(k_x)+ \sgn(k_x)c\vv{B}^\mathrm{F}_{z}(k_x)\,,
\end{align}
which is another formulation of the well known result \Eq{eqn:separation_1d_known} for electromagnetic plane waves in one dimension.

Please note that this method aims specifically at the electromagnetic fields in the absence of matter.
Including matter in the form of dielectric media or media with non-zero magnetic susceptibility by introducing the macroscopic fields $\vv{D}$ and $\vv{H}$ and adapting the method presented in this section might be possible to some degree, but is not subject of this work.

\subsection{Transversality}
In the beginning of this subsection, we assumed that the fields $\vec{E}(\vec{r}, t_0)$ and $\vec{B}(\vec{r}, t_0)$ adhere to the spectral representation introduced before.
This implies, that their Fourier transforms obey a transversality condition
\begin{align}
 \label{eqn:fourier_transversality}
 \vv{k}\cdot\vv{{E}}^\mathrm{F}({\vv{k}}) &= 0\,, & \vv{k}\cdot\vv{{B}}^\mathrm{F}({\vv{k}}) &= 0\,,
\end{align}
which will make sure that \Eqs[, ]{eqn:maxwell_gauss}{eqn:maxwell_gauss_b} and consequently \eqref{eqn:gauge_spectral} are fulfilled.
Please note that this is a necessary condition for the exact inversion of \Eq{eqn:mode_a_canonical}, which allowed us to arrive at \Eq{eqn:spectral_coefficients_a}.

This assumption, however, may be violated to some degree depending on the specifics of the numerical solver used by the simulation. Consequently, there will be some discrepancy between the reconstructed fields according to \Eqs[ and ]{eqn:kspace_final2_e}{eqn:kspace_final2_b} and the theoretical ansatz.
This discrepancy can manifest itself in the form of spurious longitudinal modes in the spectrum which can not propagate and should not exist.
In many cases, this discrepancy should be quite small and can be ignored.

To be exact, let $\vv{\mathbf{A}}_{\vv{k}}'$ with
\begin{align}
 \vv{k}\cdot\vv{\mathbf{A}}_{\vv{k}}' &\neq 0
\end{align}
represent a reconstructed vector potential that suffers from violated transversality.
Defining spectral coefficients similarly to \Eq{eqn:spectral_coefficients_a} and inserting them into \Eq{eqn:mode_a_canonical} results in
\begin{align}
 \vv{\mathbf{A}}_{\vv{k}} &= \left( \hat{e}_1 \cdot \vv{\mathbf{A}}_{\vv{k}}' \right) \hat{e}_1(\hat{k}) +\left( \hat{e}_2 \cdot \vv{\mathbf{A}}_{\vv{k}}' \right)\hat{e}_2(\hat{k})
\\ &=\vv{\mathbf{A}}_{\vv{k}}' - \hat{k}\left( \hat{k}\cdot \vv{\mathbf{A}}_{\vv{k}}' \right)\,.
\end{align}
This accounts to a projection of $\vv{\mathbf{A}}_{\vv{k}}'$ onto the space of transverse fields.

Knowing this, the spurious longitudinal modes can be removed by projecting the reconstructed spectrum onto the space spanned by the physical modes through the prescription
\begin{align}
 \vv{\mathbf{A}}_{\vv{k}} \to \vv{\mathbf{A}}_{\vv{k}} - \hat{k}\left( \hat{k}\cdot \vv{\mathbf{A}}_{\vv{k}} \right)\,.
\end{align}
Although written out here exemplarily for the vector potential, the same prescription can also be equivalently applied to $\vv{\mathbf{E}}_{\vv{k}}$ or $\vv{\mathbf{B}}_{\vv{k}}$.
Afterwards \Eqs{eqn:gauge_spectral}{eqn:maxwell_gauss_b} hold and consequently \Eq{eqn:spectral_coefficients_a}  is an exact inversion of \Eq{eqn:mode_a_canonical}.

\subsection{Temporal evolution}
\label{sec:time}
The temporal phase terms in \Eq{eqn:ansatz-a} are applicable directly to the reconstructed $\vec{k}$-space for any field component in order to evolve the field in time.
Multiplying the complex amplitudes at each wave-vector $\vec{k}$ with the temporal phase terms $e^{-\I\omega(\vec{k})\Delta t}$ will evolve fields by a time step $\Delta t$ just as they would propagate in vacuum, according to \Eqs[ and ]{eqn:ansatz_e_canonical}{eqn:ansatz_b_canonical}.
This can be used to propagate the near field data into the far field, in order to be able to compare simulation results with experimental measurements or to calculate the boundary input necessary to obtain the desired laser focus in a simulation.

\section{Numerical details}
\label{sec:numerical}
The previous section discussed the electromagnetic fields in a vacuum in the case of continuous fields and an infinite volume.
This is of course far away from numerical simulations, which need to use some kind of finite base.
In the case of finite difference time domain solvers, this finite base is characterized by a finite volume and a finite grid resolution.

\subsection{Discretization}
In the simplest case, all continuous Fourier transforms from the previous section can be simply replaced by a discrete Fourier transform.
Then this method is applicable to simulation data that is given on a homogeneous Cartesian spatial grid.
At a first glance, one could think that the finite size of the simulation domain is a large source for errors, because the discrete Fourier transform introduces only a countable set of frequency components corresponding to spatially periodic boundary conditions.
At a closer look one can reinterpret the equidistant periodic frequency components of the discrete Fourier transform as samples of a continuous spectral distribution.
This change of perspective is equivalent to padding the spatial domain with zeros that extend towards infinity.
As long as the signal that is contained in the simulation box is only supported within the box, this is a valid viewpoint and as such the finite size of the box does not induce errors.

Another concern could be the resolution of the spatial grid, possibly violating the sampling theorem.
This however is more of a problem of the underlying simulation itself, then of the Fourier transform that is applied to its output.
As all spatial frequencies that occur in the simulation had to have been properly resolved within the simulation, they can also be properly resolved in the reconstruction of the $\vv{k}$-space.

While this is not a formal proof, it is a strong hint that the accuracy and precision of this method is as such closely linked to the accuracy and precision of the underlying simulation and any errors introduced by the underlying simulation are probably larger than those introduced by discretizing the continuous Fourier transforms of our ansatz \Eq{eqn:ansatz-a}.

\subsection{Finite differences}
Apart from the deviations from the transversality conditions \Eq{eqn:fourier_transversality} which were already discussed, further discrepancies regarding \Eqs{eqn:maxwell_ampere_eqn}{eqn:maxwell_faraday_eqn} are in general to be expected from finite difference time domain Maxwell solvers.
Typically, these violations are generated by the different representation of differentiation operators in the solver and the spectral representation.

In our spectral representation, the differentiation $\frac{\partial }{\partial x}$ is represented in frequency domain by a multiplication with $\I k_x$, such that
\begin{align}
 \left( \frac{\partial F}{\partial x} \right)^\mathrm{F}(\vv{k}) = \I k_x F^\mathrm{F}(\vv{k})\,.
\end{align}
A second order accurate, central finite difference operator can also be represented in frequency domain
\begin{align}
\begin{split}
 \frac{1}{\Delta x}\left( F(x+\frac12\Delta x, y, z) -F(x-\frac12\Delta x, y, z) \right)^\mathrm{F}(\vv{k})\\
 = 2\frac{\I\sin\left( \frac{k_x}{2}\Delta x \right)}{\Delta x} F^\mathrm{F}(\vv{k})\,.
\end{split}
\end{align}
Interestingly, this difference of the representations is exactly the source of deviation of the dispersion relation on the grid $\omega_\mathrm{grid}$ with respect to the vacuum dispersion relation $\omega_0$.
Knowing this it is even more surprising, that the reconstruction of the complex $k$-space using the vacuum dispersion relation produces better results on actual simulation data, as will be demonstrated in \Sec{sec:gaussian_pulse}.

Nonetheless, as long as the wave vectors of all physically relevant contributions to the electromagnetic fields are small enough as to satisfy the small angle approximation
\begin{align}
 2\frac{\I\sin\left( \frac{k_x}{2}\Delta x \right)}{\Delta x} &\approx k_x\,,
\end{align}
these deviations should be negligable and this should be the case as long the chosen grid resolution is suitable to the physical situation.

\subsection{Staggering}
Applying the resulting \Eqs[ or ]{eqn:kspace_final2_e}{eqn:kspace_final2_b} to simulation data is straightforward, if the electric and magnetic fields are given on the same spatial grid and at the same physical time.
Unfortunately, many codes use some kind of spatial staggering and/or temporal staggering which needs to be reversed before the $\vec{k}$-space can be reconstructed.
For example, a wide-spread variant of finite difference time domain Maxwell solver is the Yee scheme \cite{Yee1966} which uses both, spatial and temporal staggering, in order to discretize Maxwell's equations with second order accurate central difference stencils, which is effectively a leapfrog method.

\subsubsection{Spatial staggering}
Spatial staggering means that different field components are defined on grids whose origins are shifted by half a grid spacing either in longitudinal or transverse direction.
In order to reverse this, the fields need to be interpolated in some way.
Having the choice between many available interpolation methods, the most natural method in the context of Fourier transforms is the application of a linear phase term $e^{\I\vv{k}\vv{r}_{\!\!\Delta}}$ to the spectral coefficients of a field component, such that the field is translated by $-\vv{r}_{\!\!\Delta}$ in the spatial domain.
In our tests on data created by the EPOCH code \cite{ArberEPOCH} which follows Yee's scheme, this method proved itself to be far superior to simple linear interpolation.

\subsubsection{Temporal staggering}
Using a leapfrog method implies, that various objects or properties thereof are updated in an alternating way.
Especially for the electromagnetic fields this typically means that the electric and magnetic field are usually not calculated at the same physical time.
Two options arise at this point.
On the one hand, one could adapt the derivation in the previous section such as to use temporally staggered fields as an input.
This should be straightforward and could be subject of a future work.
On the other hand, the temporal staggering of the fields could be removed by some kind of interpolation, similar to how the spatial staggering is removed.
In our case we opted for the second option, because many codes implicitly have routines in place, that will output the fields at equal times, even though they are internally using a temporally staggered scheme.

\subsubsection{The double leapfrog scheme}
\begin{figure}
 \includegraphics[scale=0.5]{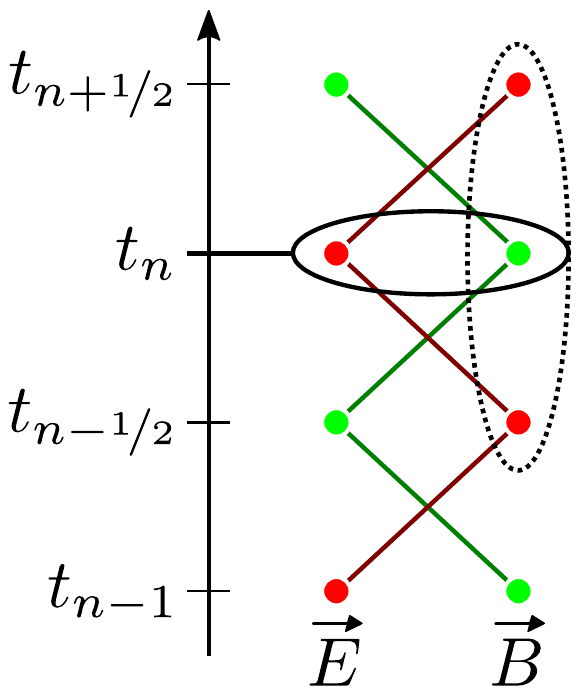}
 \caption{
 The double leap-frog scheme of some PIC codes.
 Red dots represent the fields given by the original leap-frog scheme which is only concerned with the fields that are connected with the red lines.
 Green dots represent the additional fields that are calculated by the additional half-step updates.
 They form a second group of fields, which are effectively subject to a second leap-frog scheme updating the fields with swapped roles, visualized by the green lines.
 Please note that, within the dashed ellipse, the field represented by the green dot $B(t_n)$ is given by a linear interpolation between the fields represented by the red dots $B(t_{n-\nicefrac12})$ and $B(t_{n+\nicefrac12})$.
 The fields within the solid ellipse are used to reconstruct the complex k-space at time $t$, because they are part of the same dump.
 }
 \label{fig:epoch-leapfrog}
\end{figure}
For example the EPOCH code \cite{ArberEPOCH} uses a split-step method that that produces the fields at equal times through an emerging double leapfrog scheme, see \Fig{fig:epoch-leapfrog}.
The half-step fields output by EPOCH are basically the result of a linear interpolation, as can be easily shown from EPOCH's update equations which can be found in \cite{ArberEPOCH}.
The primary leapfrog updates the magnetic field from $t=t_{n-\frac12}$ to $t=t_{n+\frac12}$ by using information from the electric field at $t=t_n$.
However, this update is split into two separate steps and both steps will apply half of that update, in order to generate the magnetic field at both times, $t=t_n$ and $t=t_{n+\frac12}$.
The magnetic field at $t=t_n$ is effectively given as the result of linearly interpolating the magnetic fields at $t=t_{n-\frac12}$ and $t=t_{n+\frac12}$ to $t=t_n$.

A linear interpolation introduces a frequency response, since linear interpolation can be viewed as a discrete convolution (represented by the operator ``$*$'': $(g*h)(n)=\sum_i x(i)*y(n+i)$) with the two element kernel $[1-a, a]$, followed by a translation.
For $0\leq a \leq 1$ we have
\begin{align*}
 g(n) &= (1-a)f(n) + af(n+1)
 \\ &= \left( [1-a, a] * f \right)(n)\,,
 \\
 \tilde{f}(n+a) &= g(n)
\\ &= (1-a)f(n) + af(n+1)\,.
\end{align*}
This operation can be represented in frequency domain by multiplying with the discrete Fourier transform of the kernel
\begin{align*}
 FT\{ [1-a, a] \} = (1-a)+a\,e^{\I \, \omega \, \Delta t}
\end{align*}
and a linear phase $e^{-\I \, \omega \, a \Delta t}$ accounting for the translation, resulting in
\begin{align}
 FT\{\tilde{f}\} = e^{-\I \, \omega \, a \Delta t} \left( (1-a)+a\,e^{\I \, \omega \, \Delta t} \right) FT\{f\}\,.
\end{align}
In case of an interpolation by a half step, $a=\tfrac12$, this simplifies to
\begin{align}
 FT\{\tilde{f}\}(\omega) = \underbrace{\cos\left( \tfrac12 \omega\,\Delta t \right)}_{=:R(\omega)} FT\{f\}(\omega)\,.
\end{align}
Please note that this frequency response completely removes waves at the Nyquist frequency $\omega_\mathrm{N} = \frac{\pi}{\Delta t}$.
This frequency response $R(\omega)$ was implicitly applied to any field that was produced by linear interpolating the fields from neighboring steps.

Performing a Fourier transformation of field data from a simulation dump from spatial domain to the frequency domain, the resulting spectrum is not a function of $\omega$, but a function of $\vv{k}$.
In order to remove the frequency response, we have to use the grid dispersion relation $\omega_\mathrm{grid}(\vv{k})$ in order to arrive at the spatial frequency response
\begin{align}
 \label{eqn:freq_resp_disp_rel}
 \tilde{R}(\vv{k}):=R(\omega_\mathrm{grid}(\vv{k}))\,.
\end{align}
Since the time-step $\Delta t$ of a PIC simulation is chosen such that all spatially resolved waves propagate with a frequency below the Nyquist frequency due to the Courant–Friedrichs–Lewy (CFL) condition \cite{York1959}, the spatial frequency response $\tilde{R}(\vv{k})$ is positive for any $\vv{k}$ present in the discrete Fourier transform of the output fields.
This means that we can divide the fields by this function after performing the discrete Fourier transformation and, with respect to the spectrum, reverse the adverse effect of the linear interpolation.
The only side effect is a slight amplification of noise.

The grid dispersion relation of simulations using Yee's scheme is given by
\begin{align*}
s_{\omega_\mathrm{grid}}^2
=s_x^2  + s_y^2+ s_z^2\label{eq:dispersion_relation}
\end{align*}
with the abbreviations
\begin{equation*}
\begin{aligned}
s_{\omega_\mathrm{grid}} &= \frac{\sin\left(\frac{1}{2}{\omega_\mathrm{grid}} \Delta t\right)}{c\Delta t}\,, \\
 s_{\{x,y,z\}} &= \frac{\sin\left(\frac{1}{2}k_{\{x,y,z\}} \Delta \{x,y,z\}\right)}{\Delta \{x,y,z\}}\,.
\end{aligned}
\end{equation*}
This can be used to find the explicit expression for the frequency response of the temporal linear interpolation, using the Yee solver dispersion relation
\begin{align}
 \label{eqn:freq_resp_lin_int_yee}
 \tilde{R}(\vv{k})
 &=  \sqrt{1-c^2\Delta t^2 \left( s_x^2 + s_y^2+ s_z^2 \right) }\,.
\end{align}

Correcting for the frequency response of the linear interpolation is now done by dividing the interpolated field (in the case displayed in \Fig{fig:epoch-leapfrog}, this is the $\vv{B}$-field), by $\tilde{R}(\vv{k})$ in frequency domain, before using \Eq{eqn:kspace_final_e}.

\section{Examples}
\label{sec:examples}
\begin{figure*}[ht]
 \includegraphics[scale=0.5]{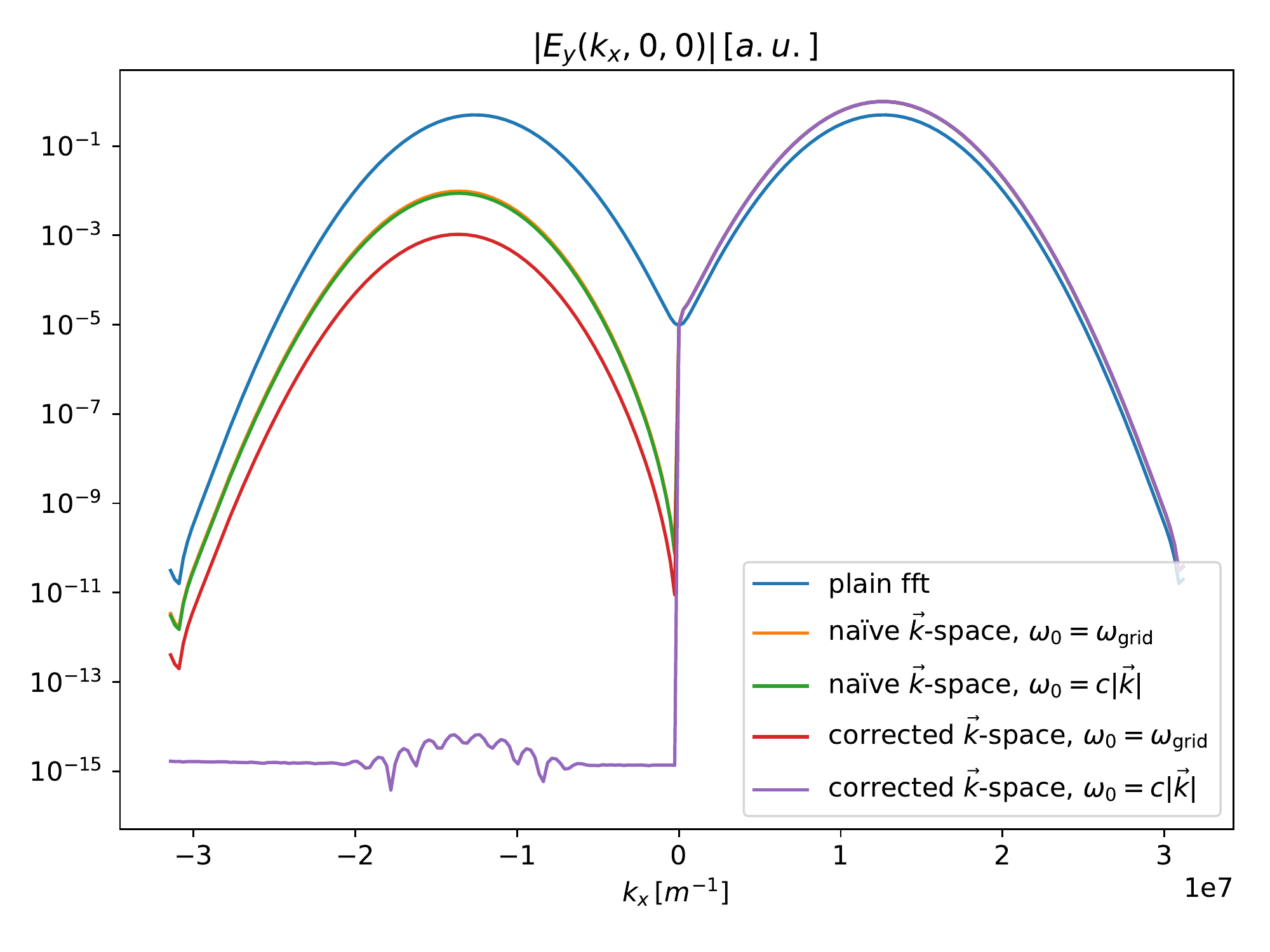}
 \hfill
 \includegraphics[scale=0.5]{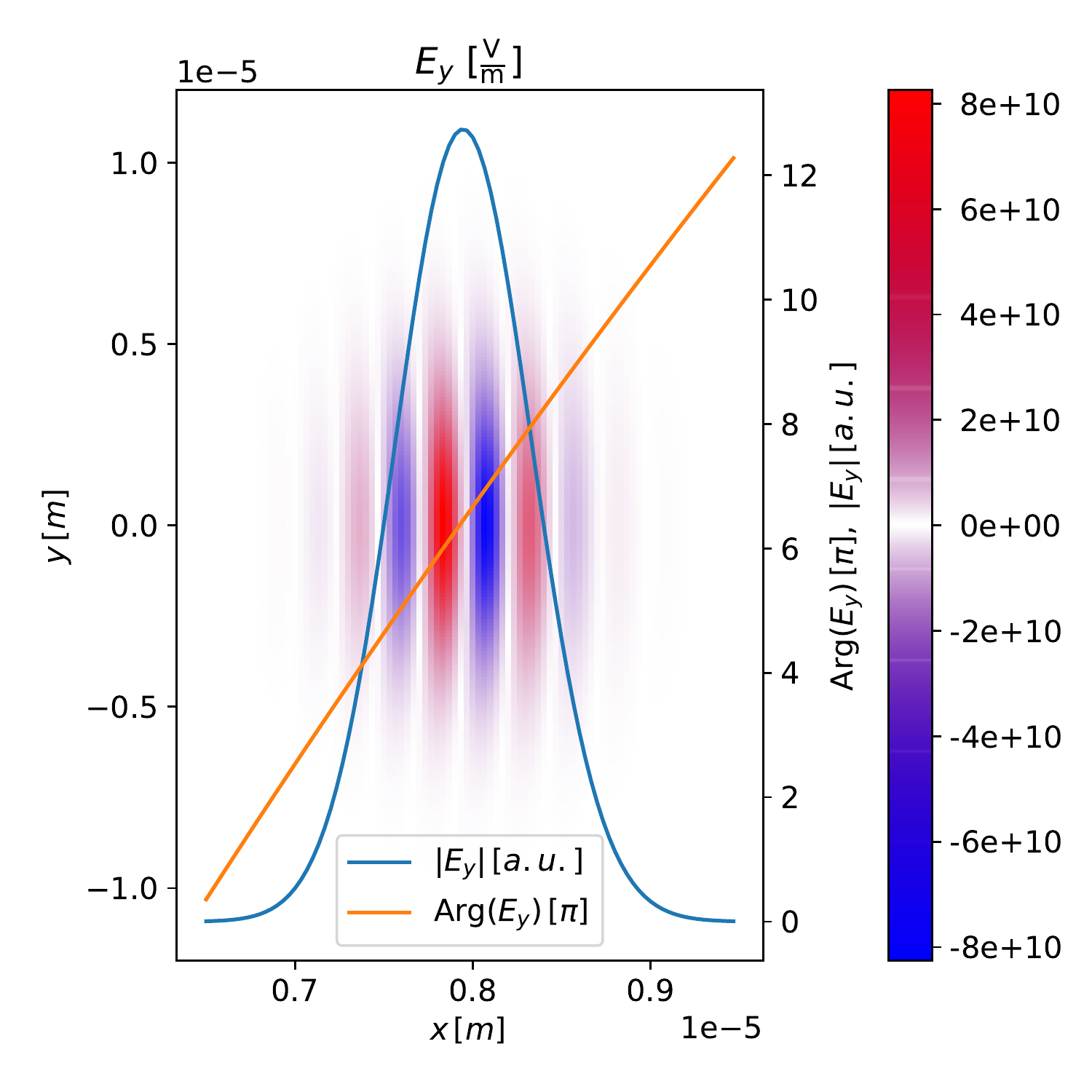}
 \caption{
 Data from a simulation containing a single Gaussian pulse moving in positive x-direction.
 Left figure: comparison between the spectra given by a simple fft, reconstructed $\protect\vv{k}$-space with (``corrected'') or without correction (``na\"ive'') for the response function of the linear interpolation (by dividing the interpolated field by $\protect\tilde{R}(\protect\vv{k})$ in frequency domain, see \Eq{eqn:freq_resp_lin_int_yee}), both combined with both options for $\omega(\protect\vv{k})$ in \Eq{eqn:kspace_final_e} each.
 The items in the figure legend are ordered according to the amplitude of the peak for negative $k_x$.
 The spectra are normalized to the maximum of the corrected $\protect\vec{k}$-space.
 All but the plain Fourier transform overlap for positive $k_x$, indicating that all results basically agree on the right-moving part of the field, the small differences are invisible due to the logarithmic scaling.
 The difference between the methods is more visible for the negative $k_x$ in that they have a different ability to demonstrate the absence of any left-moving part of the field.
 The two curves for the na\"ive method overlap, hiding the yellow curve almost completely.
 Both variants produce a left-moving ghost peak of about $10^{-2}$ of the fields amplitude.
 The method that corrects for the linear interpolation response improves this to well below $10^{-3}$ when using $\omega(\protect\vv{k})=\omega_{\mathrm{grid}}(\protect\vv{k})$.
 The corrected method using $\omega(\protect\vv{k})=ck$ is able to agree with the complete absence of any left-moving field up to 15 orders of magnitude (which is basically machine precision), shown as the purple curve.
 Right figure: envelope and phase (unwrapped using the algorithm described in \cite{PhaseUnwrap}) of the pulse along the $x$-axis as retrieved by transforming the reconstructed $\protect\vv{k}$-space back to spatial domain which yields the complex field.
 }
 \label{fig:gaussian}
\end{figure*}
The methods described in the previous sections are implemented in PostPic \cite{PostPicPaper,PostPicGitHub}, an open source python package specifically designed to aid the evaluation of data from PIC simulations.
This section will showcase the reconstruction of the physical $\vv{k}$-space from two distinct simulations performed with EPOCH\cite{ArberEPOCH}, after shortly summarizing how the reconstruction of the complex amplitudes is actually done in the case of data dumped by EPOCH.
The first example will be a single Gaussian pulse in a 3D simulation without particles, the second example is a simulation of surface high harmonic generation (SHHG) at normal incidence.

\subsection{Application of the method to EPOCH data}
\label{sec:epoch_application}
Pulling together the results from Secs.~\ref{sec:theory} and \ref{sec:numerical}, the reconstruction of the complex amplitudes of the $E_y$ component from a single data dump from EPOCH follows a number of steps.
The first step is to extract the real-valued fields $E_y(\vv{r})$, $B_x(\vv{r})$ and $B_z(\vv{r})$ from the dump, noting the positions $\vec{r}_0^{E_y}$, $\vec{r}_0^{B_x}$ and $\vec{r}_0^{B_z}$ of the origins of their respective spatially staggered grids.
After a Fourier transform is applied to each, obtaining $E^\mathrm{F}_y(\vv{k})$, $B^\mathrm{F}_x(\vv{k})$ and $B^\mathrm{F}_z(\vv{k})$, a linear phase is multiplicatively applied to two of the three fields, in order to align the grid origin points
\begin{align}
 B^\mathrm{F}_{x/z}(\vv{k}) \to e^{\I\vec{k}\cdot\left( \vec{r}_0^{E_y} - \vec{r}_0^{B_{x/z}}  \right)}B^\mathrm{F}_{x/z}(\vec{k})\,.
\end{align}
As both the electric and the magnetic field are given at the same time, one of the fields has completed a full update step, while the other is only updated half-way.
Whether this applies to the electric or the magnetic field depends on the implementation of the code and was also different for intermediate and final dumps.
In the examples presented below, the magnetic field needed to be corrected for the frequency response \Eq{eqn:freq_resp_lin_int_yee} of the half-step update according to
\begin{align}
B^\mathrm{F}_{x/z}(\vv{k})\to\frac{1}{R(\vv{k})}B^\mathrm{F}_{x/z}(\vv{k})\,.
\end{align}
After applying these two corrections, the complex amplitudes of $E_y$ can be reconstructed following \Eq{eqn:kspace_final_e}
\begin{align}
 \mathbf{E}_{y,{\vv{k}}}
&=\vv{{E}}^\mathrm{F}_y({\vv{k}}) - \frac{\omega(\vv{k})}{k^2} \left( k_x B_z^\mathrm{F}({\vv{k}}) - k_z B_x^\mathrm{F}({\vv{k}}) \right)\,.
\end{align}

\subsection{Gaussian pulse}
\label{sec:gaussian_pulse}
\begin{figure*}
 \includegraphics[scale=0.5]{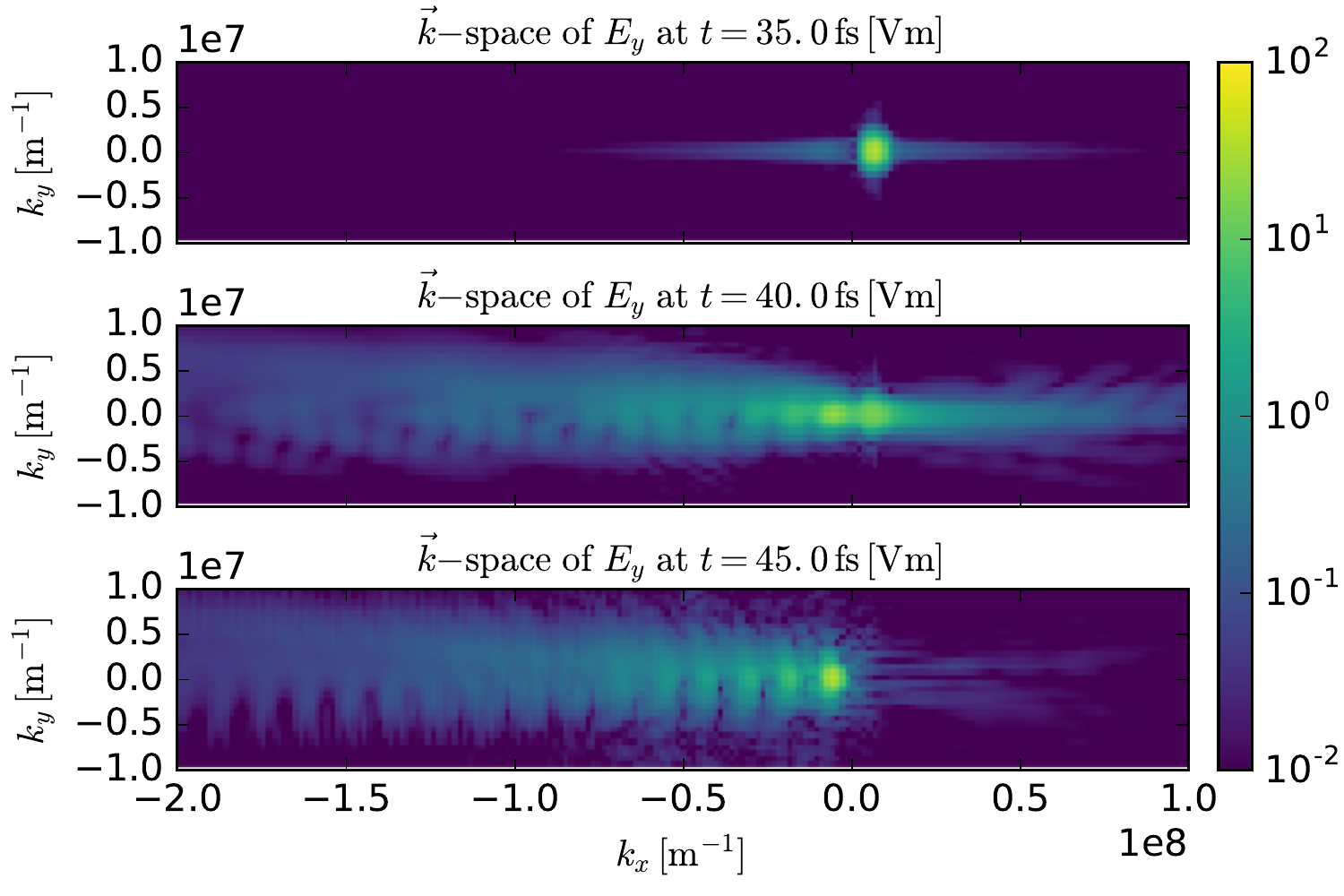}
 \hfill\hfill\hfill\hfill\hfill\hfill
 \includegraphics[scale=0.5]{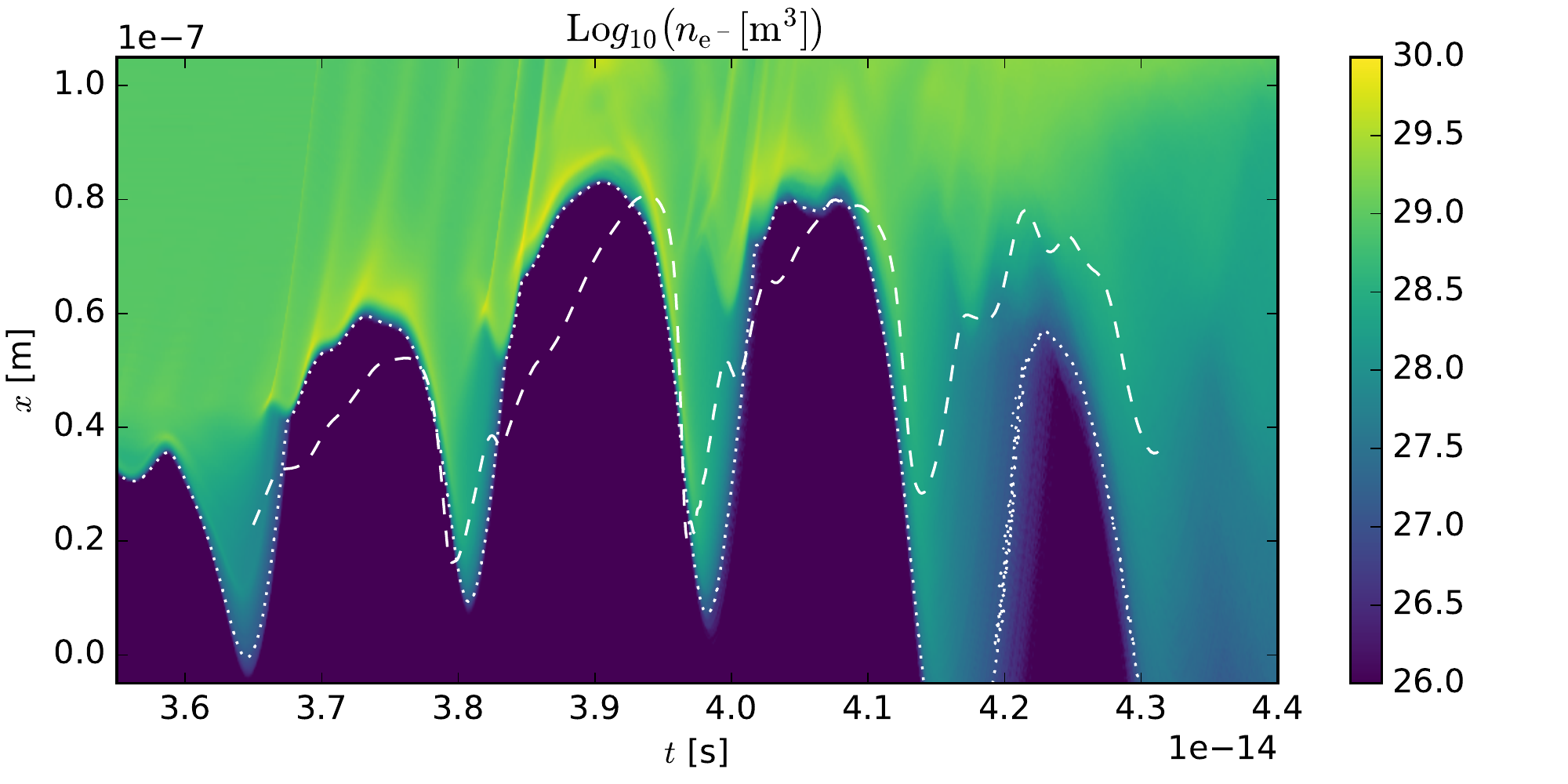}
 \caption{
 Data from a simulation of surface high harmonic generation.
 Left figure: magnitude of the reconstructed $\protect\vv{k}$-space in the beginning, middle and at the end of the interaction, from top to bottom.
 In the top figure it is clear that there is only the right-propagating laser pulse, with some residual side-lobes due to the finite extent of the used data.
 In the middle figure both are present, the leading edge of the left propagating harmonics and the tail of the incoming laser pulse.
 In the bottom figure only the left-propagating harmonics remain.
 Right figure: the color indicates the logarithm of the plasma density.
 The dotted line indicates the motion of the surface where the plasma density $n$ is equal to the critical plasma density $n_\mathrm{cr.}=\frac{\varepsilon_0 m_e \omega^2}{q_e^2}$.
 The dashed line shows the motion of the apparent reflection point as reconstructed from the longitudinal phase of the outgoing waveform according to \Eq{eqn:apparent_refl_point}, calculated from the final dump only.
 The motion of the apparent reflection point qualitatively resembles the motion of the plasma surface.
 }
 \label{fig:shhg}
\end{figure*}
This example simulation of a single Gaussian pulse in a vacuum was performed using EPOCH3D on a $720\times240\times240$ grid with a box size of $24\,\mu m$ in all directions.
A laser source was placed at the $x_\mathrm{min}$ boundary and set up with a wavelength of $\lambda_\mathrm{l}=\tfrac12\,\mu m=15\Delta x$, a beam width of $6.7\,\mu m$ FWHM and pulse duration of $3\,fs$ FWHM with its peak at $8\,fs$ after beginning the simulation.
From the grid dispersion relation we find that the laser propagates already at $\omega=0.11\,\omega_\mathrm{N}$ and is subject to a frequency response $\tilde{R}(k_\mathrm{l},0,0)=0.98$, reducing its amplitude in the interpolated field by about 2 percent.

In \Fig{fig:gaussian} the results for the reconstructed complex $\vv{k}$-space are shown.
The simple fft, shown as a blue curve, does not distinguish between the peaks at $k_\mathrm{l}$ and $-k_\mathrm{l}$ and the spectrum is symmetric.
Please note, that in \Sec{sec:theory} we had initially left open which function needs to be plugged in for $\omega(\vv{k})$ in \Eq{eqn:kspace_final_e}, presenting two options, which we both tested against this data.

If the frequency response of the linear interpolation arising from EPOCHs half-step update is not taken into account, the contrast between the forward and backward propagating peaks is around 2 orders of magnitude.
This can be understood, because the interpolation accounts for a mismatch of the electric and magnetic field amplitudes of 2 percent.
The difference between the two options for $\omega(\vv{k})$ is barely visible (orange and green curves) on the logarithmic scale.
If the frequency response is corrected, the contrast between the peaks is greatly improved.
Choosing $\omega(\vv{k})=\omega_\mathrm{grid}(\vv{k})$ leads to one additional order of magnitude of contrast (red curve), while in the case of $\omega(\vv{k})=ck$ (purple curve), a contrast of 14 orders of magnitude is achieved.
This indicates that this method has reached machine precision and that $\omega(\vv{k})=ck$ clearly is the correct choice.

When the full $\vv{k}$-space is transformed back to the spatial domain via an inverse discrete Fourier transformation, the real part is basically identical to the original, real field from the output of the simulation.
However, now the field also has an imaginary part that contains additional information, amplitude and phase.
Now information about the spectral or temporal phase of a pulse is available readily from the data.
The phase is of course wrapped to the interval $[0,2\pi]$ but unwrapping algorithms like the one described in \cite{PhaseUnwrap} may be applied.
Phase unwrapping has been applied to the phase data displayed on the right hand side of \Fig{fig:gaussian} yielding a perfectly linear phase.

\subsection{Surface High Harmonic Generation}
\begin{figure*}
 \centering
 \includegraphics[scale=0.5]{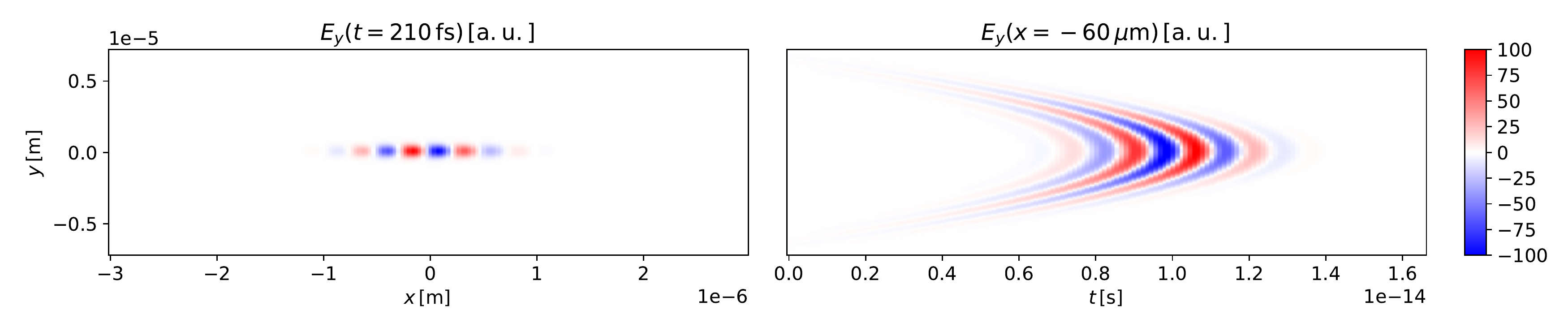}
 \caption{
    Left figure: input field data of a narrow Gaussian focus at some point in time.
    These could be actual simulation or experimental data or created using a simple analytical model for the desired focus.
    Right figure: temporal profile at a boundary $x=-60\,\mu\mathrm{m}$ that would reproduce the input field when used as input for a simulation, as computed by propagation via multiplication with a complex phase in the reconstructed $\protect\vv{k}$-space.
    The desired pulse is first propagated backwards for $210\,\mathrm{fs}$ such that it lies left of the $x=-60\,\mu\mathrm{m}$ boundary.
    After that, it is iteratively propagated forwards in steps of $dt = \tfrac{dx}{x}$, calculating the field at the boundary at each step.
 }
 \label{fig:time}
\end{figure*}
When a high intensity laser pulse is reflected at a plasma surface, the plasma surface starts to move at relativistic speeds and this motion distorts the reflected pulse.
This distortion of the pulse shape accounts for the creation of high harmonics of the lasers fudamental frequency.
The effect is called surface high harmonics generation, or SHHG for short \cite{Bulanov1994,Lichters1996,Teubner2009,Nomura2009,Dromey2009,Heissler2010,Heissler2012,Bierbach:15}.
This simulation of surface high harmonic generation at 0\textdegree{} incidence was performed with EPOCH2D.
The size of the simulation box was set to $12\,\mu m$ by $10\,\mu m$ with 800 cells per $\mu m$ in each direction.
The laser was set to a wavelength $\lambda_\mathrm{l} = 1 \, \mu m$ with a normalized intensity $a_0=20$.
The spot size of the laser in focus on the target was set to $2\,\mu m$ and the pulse duration to $3fs$ FWHM.
The plasma density was set to $n=81n_\mathrm{c}$ with an exponential ramp with a scale length $L=\tfrac{\lambda}{100}$.

Naturally, gaining insight into the generation process of the high harmonics is of high interest for such simulations. While the plasma movements can be directly observed during the interaction, the incident and outgoing electromagnetic fields are spatially superposed which usually makes them difficult to distinguish.
However, the reconstruction of the physical $\vv{k}$-space, using the method described in this paper, allows to separate the incident and outgoing waves at each time during the interaction, even when both are spatially superposed.
In this two-dimensional case, the $B_x$ field does not enter the $\vv{k}$-space reconstruction, due to $k_z=0$, see \Eq{eqn:kspace_final2_e}.
Thus, the results are based solely on the $E_y$ and $B_z$ fields.
The resulting reconstructed $\vv{k}$-spaces are displayed on the left hand side of \Fig{fig:shhg}.
From these results the left- and right-propagating waves can be separated by the sign of $k_x$.
An animation of the fields during the interaction can be found online\footnote{\url{https://youtu.be/JlygYiEQL8g}}.
\marginpar{\href{https://youtu.be/JlygYiEQL8g}{\includegraphics[width=\marginparwidth]{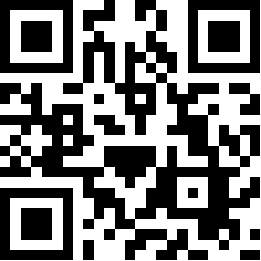}}}

Using the reconstruction of the physical k-space it is not only possible to separate incident and outgoing waves, it is also possible to gain access to the phase information of the waves.
From the phase deviation of the outgoing field from the flat phase of the incident pulse, one may infer an apparent reflection point, assuming the phase of the outgoing field is correlated with the incident pulse \cite{Geissler2007,Quere2008,Horlein2009}.
Properly defining this apparent reflection point is not completely straightforward.
In a very simple model, the deviation $\Delta\varphi$ of the phase of the outgoing field from the flat phase of the incident pulse can be used to infer the optical path length using the laser's spatial frequency $k$, resulting in
\begin{equation}
 \label{eqn:apparent_refl_point}
 \Delta x = \frac12 \frac{\Delta \varphi}{k}\,.
\end{equation}
Please note that \Eq{eqn:apparent_refl_point} is a non-relativistic approximation and allows the apparent reflection point to move faster than the speed of light.
The behaviour of the plasma surface and the apparent reflection point are displayed on the right hand side of \Fig{fig:shhg}.
Both show qualitative agreement, but a full quantitative analysis would require a better definition of the apparent reflection point that takes into account the relativistic motion of the plasma surface.
This will be subject of a future publication.
An animation of the plasma density and the apparent reflection surface during the interaction can be found online\footnote{\url{https://youtu.be/U0RmoGLrgDk}}.
\marginpar{\href{https://youtu.be/U0RmoGLrgDk}{\includegraphics[width=\marginparwidth]{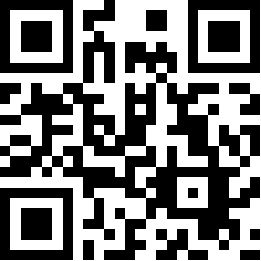}}}

\subsection{Calculating the far field}
Using temporal evolution in Fourier space as briefly discussed in \Sec{sec:time}, it is possible to calculate a temporal field profile, e.\,g. $E_y(t, y, z)$, that would hit a detector located at a given plane in the far field, e.\,g. defined by $x=0$.
This is in fact the exact same task as the calculation of the boundary input necessary to achieve the desired focus field in a simulation.
In order to do this, an initial (possibly negative) $\Delta t$ is applied to the complex $\vv{k}$-space data $\mathbf{E}_{y,\vec{k}}$, such that the pulse lies just in front of the surface at which we would like to record the temporal profile.
After that, the pulse is propagated iteratively in small steps $dt=\tfrac{dx}{c}$ until it passed completely through the plane in question.

At each step, the $k_x$ axis is transformed back into the spatial domain, calculating only the values for $x=0$.
Because only the plane  $x=0$ is of interest in this operation, and at $x=0$ the exponential $e^{\I k_x x}$ becomes unity, instead of performing a full Fourier transform it is sufficient to perform a simple integral along the $k_x$.
By globally shifting the coordinates of the $x$-axis, this method can also be used for planes defined by non-vanishing values of a coordinate.

The resulting two dimensional fields are collected, to obtain a temporal profile $E_y(t, k_y, k_z)$.
A 2D inverse Fourier transform of the $(k_y, k_z)$-plane may be optionally computed afterwards to get the actual field $E_y(t, y, z)$ in spatial domain.
An example of this is depicted in \Fig{fig:time}.

\section{Conclusion}
The presented method is able to accurately reconstruct the physical $\vv{k}$-space including its phase information from real-valued field data from simulations.
Compared to earlier approaches, it requires only a single snapshot that contains the real-valued electromagnetic fields, which is part of the standard output for most kinds of simulations.
This makes this method easy to use and apply even to already existing simulation data. Our approach is also computationally fast as it requires only three Fourier transforms to directly compute the complex $\vv{k}$-space.

A ready-to-use implementation is available as part of PostPic, an easy-to-use python package suitable for evaluation of many kinds of particle in cell simulation output files.
This will hopefully lead to a more widespread use of $\vv{k}$-space analysis in various fields of physics that deal with simulations that contain electromagnetic fields, allowing easy access to spatial and temporal phases of all propagating waves in the simulation, even when counter propagating waves are spatially superposed.

Beyond the reconstruction described in this work, the complex $\vv{k}$-space is the prerequisite for various kinds of further analysis, such as phase retrieval of the EM field or vacuum field propagation over arbitrary distances.
This provides access to far field data that can be directly compared with data collected by experiments, as well as supplying the necessary input data for simulations to recreate any kind of focal field configuration.

\end{document}